\documentclass[11pt]{article}
\usepackage[textwidth=15.2cm,textheight=22cm]{geometry}
\usepackage{amsmath,amssymb}
\usepackage{latexsym}
\usepackage{multicol}
\usepackage{graphicx}
\usepackage{bm}
\allowdisplaybreaks[1]

\newcommand{\del}{\partial}
\newcommand{\be}{\begin{equation}}
\newcommand{\ee}{\end{equation}}
\newcommand{\ba}{\begin{eqnarray}}
\newcommand{\ea}{\end{eqnarray}}
\newcommand{\bdm}{\begin{displaymath}}
\newcommand{\edm}{\end{displaymath}}

\def\ba{\bar A}

\def\beq{\begin{equation}}
\def\eeq{\end{equation}}
\newcommand{\half}{\frac{1}{2}}
\newcommand{\nn}{\nonumber}

\newcommand{\ndt}{\noindent}

\newcommand{\nbar}[1]{\overline{#1}}

\def\bea{\begin{eqnarray}}
\def\eea{\end{eqnarray}}
\def\beas{\begin{eqnarray*}}
\def\eeas{\end{eqnarray*}}
\def\sla{\raise.15ex\hbox{$/$}\kern-.57em}

\def\parm{{\partial}_{-}}
\def\parp{\partial^+}

\def\spa#1.#2{\left\langle#1\,#2\right\rangle}
\def\spb#1.#2{\left[#1\,#2\right]}

\begin{document}

\begin{titlepage}
\begin{flushright}    
{\small $\,$}
\end{flushright}
\vskip 1cm
\centerline{\Large{\bf{Deriving field theories for particles of arbitrary spin}}}
\vskip 0.5cm
\centerline{\Large{\bf{with and without supersymmetry}}}
\vskip 1.5cm
\centerline{Sudarshan Ananth}
\vskip .5cm
\centerline{\it {Indian Institute of Science Education and Research}}
\centerline{\it {Pune 411008, India}}
\vskip 1.5cm
\centerline{\bf {Abstract}}
\vskip 0.2cm
\ndt We review the derivation of light-cone interaction vertices for fermionic and bosonic fields of arbitrary spin. The resulting amplitudes and their factorization properties are discussed. We then show how this symmetry-based approach works for theories with extended supersymmetry like $\mathcal N=4$ Yang-Mills theory and $\mathcal N=8$ supergravity. 
\end{titlepage}

\section{Introduction}

\ndt At the level of the equations of motion, there has been considerable progress in our understanding of higher spin theories~\cite{MV}. However, a description of higher spin fields ($\lambda>2$) in a Lagrangian formalism, essential to quantization, remains elusive. There are various no-go theorems relating to higher spin fields and one of the things they tell us is that we cannot write down a consistent interacting theory of massless higher spin fields in flat spacetime. In contradiction to this statement is the result of~\cite{BBB} where a consistent cubic interaction vertex was constructed in the light-cone formalism. Given that the cubic interaction vertex is in some sense trivial and may be written down based on helicity and dimensional considerations~\cite{smat} the question then becomes whether a quartic interaction vertex can be arrived at within the same formalism. The discussions in~\cite{smat} outline why a BCFW~\cite{BCFW} type construction is bound to fail but this does not rise to the level of a proof. It is worth noting, at this stage, that most of the no-go literature pertaining to higher spin theories~\cite{BBS} assumed theories with manifest locality and manifest Lorentz invariance. Neither of these properties is manifest in light-cone gauge and this is one motivation to study higher spin fields in this gauge. Another motivation is the exclusive focus on the physical degrees of freedom. Since the first paper on light-cone cubic interaction vertices, there has been considerable work on the subject~\cite{RRM,bengt,AA,SA}.
\vskip 0.3cm
\ndt The procedure we review here, is that initiated in~\cite{BBB}. We revisited their study~\cite{SA} in the light of recent advances in field theory~\cite{dixon}. These new methods, when introduced to the older analysis of~\cite{BBB} yielded among other results, a Lagrangian origin~\cite{AT} for the KLT relations~\cite{KLT} and factorization properties~\cite{AA}. The formalism in~\cite{BBB} starts with just the Poincar\'e algbera, for four dimensional flat spacetime, in light-cone gauge. Among the generators of this algebra is $+$ component of momentum which is the light-cone Hamiltonian. An ansatz for this Hamiltonian is made based on dimensional analysis and helicity and then refined by requiring closure of the Poincar\'e algebra\footnote{In light-cone gauge, Poincar\'e invariance needs to be checked as the formalism is not manifestly covariant.}. This leads to a class of higher spin cubic vertices.
\vskip 0.3cm
\ndt This framework is particularly interesting in the context of aribtrary spin theories in non-flat spacetime backgrounds. The formalism described here may be extended to $AdS_4$ (for the spin$=\!\!2$ case, see~\cite{AAM}). The cubic interaction vertices are straightforward to derive in this background but the quartic interaction vertices are challenging. This approach is likely to yield key ingredients necessary to establish a Lagrangian origin to the Vasiliev program~\cite{MV}. In other words, a derivation of consistent quartic interaction vertices involving higher spin fields in $AdS_4$ should necessitate the inclusion of a tower of higher spin fields thus providing a Lagrangian origin to the Vasiliev equations of motion.
\vskip 0.3cm
\ndt The approach we review is particularly powerful when applied to supersymmetric theories where we have more than just the Poincar\'e algebra to work with. Supergravity in eleven dimensions~\cite{Julia}, when reduced to $d=4$, becomes the maximally supersymmetric ${\cal N}=8$ theory. Its light-cone formulation in superspace  is incomplete because its higher-point interactions are not known in terms of a light-cone superfield. A simpler, yet in many ways similar, theory in $d=4$ is the maximally supersymmetric ${\cal N}=4$ Yang-Mills theory. The ${\cal N}=8$ and ${\cal N}=4$ theories are each described by a single light-cone superfield which captures their physical degrees of freedom. Both theories may be oxidized~\cite{ABR} to their higher-dimensional parent theories, yielding superspace descriptions without auxiliary fields. However, while there exist two distinct algebraic methods to derive the entire classical Hamiltonian of $\mathcal N=4$ Yang-Mills there is considerable difficulty in deriving the $\mathcal N=8$ theory even to quartic order.

\vskip 0.3cm
\section{Poincar\'e generators}

\ndt We define light-cone co-ordinates in $(-,+,+,+)$ Minkowski space-time by
\begin{eqnarray}
x^{\pm}=\frac{x^{0}\pm x^{3}}{\sqrt{2}} \;,\qquad
x = \frac{x^{1}+ix^{2}}{\sqrt{2}} \;,\qquad\bar{x}= \frac{x^{1}-ix^{2}}{\sqrt{2}}\ .
\end{eqnarray}
The corresponding derivatives are $\partial_{\pm}\,,\,\,\bar{\partial}$ and $\partial$. One of the reasons $d=4$ is so special is that {\it {all}} massless fields have exactly two physical degrees of freedom, which we call $\phi$ and $\bar{\phi}$. The field $\phi$ has helicity $\lambda$ while the field $\bar\phi$ has helicity $-\lambda$. The light-cone generators of the Poincar\'{e} algebra are
\bea
p^{-}=i\frac{\partial\bar{\partial}}{\partial_{-}}=-p_+ \qquad p^+=-i\partial^{+}=-p_- \qquad  \bar{p}=-i\bar{\partial}\ ,
\eea
\begin{eqnarray}
j\!\!\!\!\!\!&= i(x\bar{\partial}-\bar{x}\partial - \lambda) \;,\qquad  &j^{+} = (x^+ \partial-x\partial^+)\ , \nn\\
 j^{+-}\!\!\!\!&=(x^{+}\frac{\partial\bar{\partial}}{\partial^+}-x^-\partial^+ ) \;,\qquad & j^-=(x^{-}\partial-x\frac{\partial\bar{\partial}}{\partial^+}+\lambda \frac{\partial}{\partial^+}) \ ,
\end{eqnarray} 
and their complex conjugates with $\frac{1}{\parm}$ defined following~\cite{SM}. For the free theory, $\partial_{+}=\frac{\partial\bar{\partial}}{\partial_{-}}$ and the Hamiltonian reads
\begin{equation}
H\equiv\int d^{3}x\,\mathcal{H}=-\int d^3x\,\bar\phi\,\partial\bar\partial\,\phi\  ,
\end{equation}
also written as
\begin{equation}
\label{hamilx}
H\equiv\int d^{3}x\,\mathcal{H}=\int d^{3}x\,\partial_{-}\bar{\phi}\,\delta_{\mathcal{H}}\phi
\ ,
\end{equation}
where the time translation operator is introduced through the Poisson bracket
\begin{eqnarray}
\delta_{\mathcal{H}}\phi\equiv\partial_{+}\phi=\lbrace \phi,\mathcal{H}\rbrace\ .
\end{eqnarray}
\ndt On the light-cone, spacetime symmetries split into two types. Kinematical symmetries are unaltered by interactions while dynamical symmetries pick up corrections and are {\em non-linearly} realized on the fields. In supersymmetric theories, the supersymmetries also separate into dynamical and kinematical supersymmetries. For the interacting theory, $\delta_{\mathcal H}$ picks up corrections, order by order in the coupling constant $\alpha$, as do
\bea
\delta_{j^{+-}}\phi\;\;; \quad \delta_{j^{-}}\phi\;\; ;\quad \delta_{\bar{j}^{-}}\phi\ .
\eea
These corrections, non-linear in nature, need to be constructed. 

\section{Bosonic fields}
\vskip 0.2cm
\ndt We review here, the procedure to derive cubic interaction vertices for three bosonic fields of arbitrary spin. For additional details, we refer the reader to~\cite{AA}. At cubic order, the following structures (at order $\alpha$) appear in the Hamiltonian
\bea
\delta^{\alpha}_{\mathcal{H}}\phi_1\sim\phi_2\phi_3\ ; \quad \delta^{\alpha}_{\mathcal{H}}\phi_2\sim\phi_1\phi_3\ ;\quad \delta^{\alpha}_{\mathcal{H}}\phi_3\sim\phi_1\phi_2\ .
\eea
The fields $\phi_1$, $\phi_2$ and $\phi_3$ have integer spins $\lambda_1$, $\lambda_2$ and $\lambda_3$ respectively. In terms of the action, the first structure would correspond to 
\bea
{\mbox {\it {S}}}\sim\int \,d^4x\,\;\bar\phi_1\,\phi_2\,\phi_3\ +c.c.\;.
\eea
We sprinkle in derivatives to arrive at the ansatz 
\begin{equation}
\label{integers}
\delta^{\alpha}_{\mathcal{H}}\phi_1=\alpha\,A\,\partial^{+\mu}\left[\bar\partial^a\partial^{+\rho}\phi_2\bar\partial^b\partial^{+\sigma}\phi_3\right]\ +c.c.\;.
\end{equation}
$\mu ,\rho ,\sigma ,a,b$ are integers and $A$ is a number. The commutators
 \begin{eqnarray}
&&\left[\delta_{j},\delta_{\mathcal{H}}^{\alpha}\right]\phi_1 =0\ , \nn \\
&&\left[\delta_{j^{+-}},\delta_{\mathcal{H}}^{\alpha}\right]\phi_1 =-\delta_{\mathcal{H}}\phi_1\ ,
\end{eqnarray}
imply that
\begin{eqnarray}
\label{conds1}
\nn a+b\!\!\!\!\!\!&&=\lambda_2+\lambda_3-\lambda_1  \\
\mu +\rho +\sigma\!\!\!\!\!\!&&=-1 \ .
\end{eqnarray} 
As $a,b>0$, the first condition implies that the vertex exists only if $\lambda_2+\lambda_3>\lambda_1$. Let $\lambda \equiv \lambda_2 +\lambda_3 -\lambda_1$ so $a+b = \lambda$. The ansatz is now a sum of the $(\lambda + 1)$ terms
\be
\label{ansatz}
\delta_H^\alpha \phi_1=\alpha\sum_{n=0}^{\lambda}A_n\,\partial^{+\mu_n}\left[\bar\partial^n\partial^{+\rho_n}\phi_2\,\bar\partial^{(\lambda -n)}\partial^{+\sigma_n}\phi_3\right]\ +c.c.\;.
\ee
\\
\ndt The commutators 
\bea
\left[\delta_{\bar{j}^{-}}, \delta_{H}\right]^{\alpha}\phi_{1} =0 \qquad \left[\delta_{j^{+}}, \delta_{H}\right]^{\alpha}\phi_{1} = 0\;,
\eea
yield the conditions
\bea
\label{conds3}
\nn \sum_{n=0}^{\lambda}\!\!\!\!\!\!&& A_n\,{\biggl \{}\;(\mu_n +1-\lambda_1)\del^{+(\mu_n -1)}\bar{\del}(\bar{\del}^{n}\del^{+\rho_n}\phi_2\bar{\del}^{(\lambda -n)}\del^{+\sigma_n}\phi_3)\\
&&+(\rho_n +\lambda_2)\del^{+\mu_n} (\bar{\del}^{(n+1)}\del^{+(\rho_n -1)}\phi_2\bar{\del}^{(\lambda -n)}\del^{+\sigma_n}\phi_3)\\	
\nn &&+(\sigma_n +\lambda_3)\del^{+\mu_n} (\bar{\del}^n\del^{+\rho_n}\phi_2\bar{\del}^{(\lambda-n+1)}\del^{+(\sigma_n -1)}\phi_3){\biggr \}}\ =0\ ,\\
\nn\sum_{n=0}^{\lambda}\!\!\!\!\!\!&& A_n{\biggl \{}\;n\,\del^{+\mu_n}(\bar{\del}^{(n-1)}\del^{+(\rho_n +1)}\phi_{2}\bar{\del}^{(\lambda -n)}\del^{+\sigma_n}\phi_{3})\\&&+(\lambda -n)\del^{+\mu_n}(\bar{\del}^{n}\del^{+\rho_n}\phi_{2}\bar{\del}^{(\lambda -n-1)}\del^{+(\sigma_n +1)}\phi_{3}){\biggr \}}=0\ ,
\eea
which are satisfied when the coefficients obey
\be
\label{A}
\nn A_{n+1}=-\frac{(\lambda - n)}{(n+1)}A_n \,\,=\,\, (-1)^{(n+1)} \binom{\lambda}{n+1}A_0\ ,
\ee
\be
\label{rho}
\rho_{n+1}=\rho_n - 1 \quad;\quad \sigma_{n+1}=\sigma_n +1 \quad ;\quad \mu_{n+1}=\mu_n \: ,
\ee
with the conditions
\be
\label{boundary}
\rho_{{\,}_{n=\lambda}}=-\lambda_2 \qquad \sigma_{{\,}_{n=0}}= -\lambda_3.
\ee
We find 
\be
\rho_n = \lambda - \lambda_2 -n \ ; \qquad \sigma_n = n-\lambda_3 \ ;\qquad \mu_n=\lambda_1-1\ .
\ee
\\
Thus (\ref{ansatz}) becomes
\be
\delta_H^\alpha \phi_1=\alpha\sum_{n=0}^{\lambda}(-1)^n \binom{\lambda}{n}\,\partial^{+(\lambda_1 -1)}\left[\bar\partial^n\partial^{+(\lambda -\lambda_2-n)}\phi_2\,\bar\partial^{(\lambda -n)}\partial^{+(n-\lambda_3)}\phi_3\right]+c.c.\: .
\ee
\\
Using
\be
H = \int d^{3}x\,\,\partial_{-}\bar{\phi}_1\,\delta_{\mathcal{H}}\phi_1,
\ee
the interacting Hamiltonian is
\bea
\label{hamilt}
H^{\alpha}=\alpha \int d^{3}x\,\,  \sum_{n=0}^{\lambda}(-1)^n \binom{\lambda}{n}\,\bar{\phi}_1\,\partial^{+\lambda_1}\left[\bar\partial^n\partial^{+(\lambda -\lambda_2 -n)}\phi_2\,\bar\partial^{(\lambda -n)}\partial^{+(n-\lambda_3)}\phi_3\right]+c.c.\: .
\eea
\\
The power of this approach is clear. One example: $H^\alpha$ vanishes for $\lambda_1=\lambda_2=\lambda_3=\lambda '$ in (\ref{hamilt}) for odd $\lambda '$ making the introduction of a gauge group mandatory for odd integer spins. 

\vskip 0.3cm

\subsection{Factorization and perturbative ties} 
\vskip 0.1cm
\ndt The above results may be written in the language of spinor helicity products~\cite{dixon}
\begin{equation}
<kl>= \sqrt{2}\:\frac{(kl_{-}-lk_{-})}{\sqrt{k_{-}l_{-}}} \qquad  [kl]= \sqrt{2}\:\frac{(\bar{k}l_{-}-\bar{l}k_{-})}{\sqrt{k_{-}l_{-}}}\ .
\end{equation}
\\ 
Equation (\ref{hamilt}) contains two kinds of terms: $\!\bar\phi \phi \phi$ and $\phi \bar\phi\bar\phi$ and the Fourier coefficient of the second variety $\phi_1(p)\bar\phi_2(k)\bar\phi_3(l) \: \delta^{4}(p\!+\!k\!+\!l)$  is
\bea
\frac{p_{-}^{\lambda_1}}{k_{-}^{\lambda_2}l_{-}^{\lambda_3}}
(lk_--l_-k)^{\lambda_2+\lambda_3-\lambda_1},
\eea
\\
which simplifies to
\bea
\label{finala}
\frac{1}{\sqrt{2^{\lambda}}}\,\,\,<pk>^{^{(-\lambda_1 +\lambda_2 -\lambda_3)}}\:\:<kl>^{^{(\lambda_1 +\lambda_2 +\lambda_3)}}\:\:<lp>^{^{(-\lambda_1 -\lambda_2 +\lambda_3)}}.
\eea
\vspace{.1cm}
\\
This means that given vertices for $(\lambda_1, \lambda_2, \lambda_3)$ and $(\lambda '_1, \lambda '_2, \lambda '_3)$, their product yields the vertex for $(\lambda_1 + \lambda '_1, \lambda_2 + \lambda '_2, \lambda_3 + \lambda '_3)$. Further, the coefficient for $(n\lambda_1 ,n\lambda_2 , n\lambda_3 )$ is that for $(\lambda_1 ,\lambda_2 , \lambda_3 )$ raised to the power $n$~\cite{AA}. In particular, $\lambda_1=\lambda_2=\lambda_3=1$ and $\lambda_1=\lambda_2=\lambda_3=2$ offers a field-theoretic realization~\cite{AT} of the KLT relations~\cite{KLT}.
\vskip 0.3cm

\section{Fermionic fields}
\vskip 0.2cm
\ndt We now introduce fermions into this formalism so we can construct interactions involving matter. The $4\times 4$ Gamma matrices are
\bea
\gamma^0 = \left(\begin{array}{cc}
0 & 1 \\
1 & 0 \end{array}\right)
\hspace{1cm}\gamma^i = \left(\begin{array}{cc}
0 & \sigma^i \\
-\sigma^i & 0 \end{array}\right)
\hspace{1cm}C = \left(\begin{array}{cc}
i\sigma^2 & 0 \\
0 & -i\sigma^2 \end{array}\right)\ ,
\eea
and we define
\bea
&&P_+ \equiv \half \gamma_+\gamma_- \\
&&P_- \equiv \half \gamma_-\gamma_+\ ,
\eea
where
\bea
\nn
&&\gamma_+ = \frac{1}{\sqrt{2}}\left(\gamma_0+\gamma_3\right) \\
&&\gamma_- = \frac{1}{\sqrt{2}}\left(\gamma_0-\gamma_3\right).
\eea
To a Dirac spinor with Grassmann valued components
\be
\Psi = \left(\begin{array}{c}
\chi_1\\
\chi_2\\
\chi_3\\
\chi_4 \end{array}\right)\ .
\ee
we apply the Majorana condition $\Psi = C\bar{\Psi}^T$ to find
\bea
\label{maj}
\chi_1 = \bar{\chi}_4 \qquad\ ; \qquad \chi_2 = -\bar{\chi}_3\ .
\eea
We introduce
\bea
\Psi_+ = P_+\Psi \hspace{0.5cm};\hspace{0.5cm} \Psi_- = P_-\Psi\ ,
\eea
which satisfy
\bea
\label{kin}
\del_-\Psi_- = \half \gamma_-\gamma_i\del_i\Psi_+ \qquad\qquad i = 1, 2\; ,
\eea
\bea
\label{dyn}
\del_+\Psi_+ = \half \gamma_+\gamma_i\del_i\Psi_- \qquad\qquad i = 1, 2\; .
\eea
The first (kinematic) relation yields
\be
\label{plus}
\chi_1 = \frac{\bar{\del}}{\del_-}\chi_2\ .
\ee
The entire spinor, in terms of $\chi_3 \equiv \chi$, is
\be
\label{reduced}
\Psi = \left(\begin{array}{c}
-\frac{\bar{\del}}{\del_-}\bar{\chi} \\
-\bar{\chi}\\
\chi\\
-\frac{\del}{\del_-}\chi \end{array}\right)\ .
\ee
The free light-cone action for a fermionic field of half-integer spin $\lambda$ is
\bea
\label{fermion}
S = \int d^{4}x\,\,\, i \bar\psi\,\frac{\Box}{\del_-}\,\psi .
\eea
The Poincar\'e generators are
\bea
\delta_{j}\!\!\!\!\!\!&\psi= i(x\bar{\partial}-\bar{x}\partial + \lambda)\psi \;,\qquad  &\delta_{j^{+}}\psi = (x^+ \partial-x\partial^+)\psi\;, \nn\\
\delta_{j^{+-}}\,\,\,\!\!\!\!\psi &=(x^{+}\frac{\partial\bar{\partial}}{\partial^+}-x^-\partial^+ -\half)\psi \;,\qquad & \delta_{j^-}\psi=(x^{-}\partial-x\frac{\partial\bar{\partial}}{\partial^+}+(\lambda +\half) \frac{\partial}{\partial^+})\psi\; , \
\eea
and their conjugates. Unlike the bosonic case, $\bar{\psi}$ has helicity $\lambda$, a positive half integer.

\vskip 0.3cm
\subsection{Interacting arbitrary spin fields} 
\vskip 0.2cm

\ndt The free Hamiltonian involving a massless spin $\lambda$ boson and a massless spin $\lambda'$ fermion is
\begin{equation}
\label{hamil}
H = \int d^3x\,\,\,\left(-\bar\phi\,\partial\bar\partial\,\phi\ +i \bar\psi\,\frac{\partial\bar\partial}{\del_-}\,\psi\right)=\int d^{3}x\,\,\left(\partial_{-}\bar{\phi}\,\delta_{\mathcal{H}}\phi + i \bar\psi\,\delta_{\mathcal{H}}\,\psi\right)\ .
\end{equation}
We start with an ansatz for the interactions from $\delta_{\mathcal{H}}^\alpha \phi_1$ having the structure $\bar{\phi}_1\psi_2\psi_3$ where the fields $\phi_1$, $\psi_2$ and $\psi_3$ carry spins $\lambda_1$ (integer), $\lambda_2$ and $\lambda_3$ (both half integers) respectively. For a detailed description of the following, we refer the reader to~\cite{AA2}. The ansatz reads
\begin{equation}
\label{integers}
\delta^{\alpha}_{\mathcal{H}}\phi_1=\alpha\,A\,\partial^{+\mu}\left[\partial^a\partial^{+\sigma}\psi_2\partial^b\partial^{+\rho}\psi_3\right]\  .
\end{equation}
Again, $\mu ,\rho ,\sigma ,a,b$ are integers and $A$ is a number. To ensure Lorentz invariance, we demand closure of the Poincar\'e  algebra to this order in $\alpha$. The kinematical constraints
\begin{eqnarray}
&&\left[\delta_{j},\delta_{\mathcal{H}}^{\alpha}\right]\phi_1 =0\ \hspace{1cm}\left[\delta_{j^{+-}},\delta_{\mathcal{H}}\right]^{\alpha}\phi_1\nn =-\delta^\alpha_{\mathcal{H}}\phi_1\ ,
\end{eqnarray}
impose the constraints
\begin{eqnarray}
\label{conds1}
\nn a+b\!\!\!\!\!\!&&=\lambda_1+\lambda_2+\lambda_3\equiv \lambda  \\
\mu +\rho +\sigma\!\!\!\!\!\!&&=-2 \ .
\end{eqnarray} 
While
\bea
\left[\delta_j^{-}, \delta_{H}\right]^{\alpha}\phi_{1} =0 \qquad \left[\delta_{\bar j^{+}}, \delta_{H}\right]^{\alpha}\phi_{1} = 0\: ,
\eea
require that
\bea
\label{conds3}
\nn \sum_{n=0}^{\lambda}\!\!\!\!\!\!&& A_n\,{\biggl \{}\;-(\mu_n +1+\lambda_1)\del^{+(\mu_n -1)}\del(\del^n\del^{+\rho_n}\psi_2\del^{(\lambda -n)}\del^{+\sigma_n}\psi_3)\\
&&-(\sigma_n +\lambda_2+\half)\del^{+\mu_n} (\del^{(n+1)}\del^{+(\sigma_n -1)}\psi_2 \del^{(\lambda -n)}\del^{+\rho_n}\psi_3)\\	
\nn &&-(\rho_n +\lambda_3+\half)\del^{+\mu_n} (\del^n\del^{+\sigma_n}\psi_2 \del^{(\lambda-n+1)}\del^{+(\rho_n -1)}\psi_3){\biggr \}}\ =0\ ,\\
\nn\sum_{n=0}^{\lambda}\!\!\!\!\!\!&& A_n{\biggl \{}\;n\,\del^{+\mu_n}(\del^{(n-1)}\del^{+(\sigma_n +1)}\psi_{2} \del^{(\lambda -n)}\del^{+\rho_n}\psi_{3})\\&&+(\lambda -n)\del^{+\mu_n}(\del^{n}\del^{+\sigma_n}\psi_{2}\bar{\del}^{(\lambda -n-1)}\del^{+(\rho_n +1)}\psi_{3}){\biggr \}}=0\ .
\eea
\\
We find
\be
\label{A}
\nn A_{n}= (-1)^{(n)} \binom{\lambda}{n};\qquad \sigma_n = -\lambda - \lambda_2 +n-\half \ ; \qquad \rho_n = n-(\lambda_3+\half) \ ;\qquad \mu_n=-(\lambda_1+1)\ ,
\ee
\\
so (\ref{integers}) becomes
\be
\delta_H^\alpha \phi_1=\alpha\sum_{n=0}^{\lambda}(-1)^n \binom{\lambda}{n}\,\frac{1}{\partial^{+(\lambda_1 +1)}}\left[\partial^n\frac{\partial^{+(\lambda-n)}}{\partial^{+(\lambda_2+\half)}}\psi_2\,\partial^{(\lambda -n)}\frac{\partial^{+n}}{\partial^{+(\lambda_3+\half)}}\psi_3\right]\: .
\ee
\vskip 0.2cm
\ndt From which we obtain the Hamiltonian and thus the action 

\bea
\label{Action}
S = \int d^4x\hspace{-.5cm} &&\left[\half\bar\phi_1\Box\phi_2 + i \bar\psi_2\,\frac{\Box}{\del_-}\,\psi_2 + i \bar\psi_3\,\frac{\Box}{\del_-}\,\psi_3\right.\\
\nn 
&&\left. + \,\alpha \bar{\phi}_1\sum_{n=0}^{\lambda}(-1)^n \binom{\lambda}{n}\,\frac{1}{\partial^{+\lambda_1}}\left[\partial^n\frac{\partial^{+(\lambda-n)}}{\partial^{+(\lambda_2+\half)}}\psi_2\,\partial^{(\lambda -n)}\frac{\partial^{+n}}{\partial^{+(\lambda_3+\half)}}\psi_3\right]\right]\ .
\eea
\vskip 0.2cm
\ndt Once again, the algebra teaches us that for odd helicity $\phi_1$, self-interactions require an internal symmetry group with an antisymmetric structure constant. This permits two varieties of vertices: $t_a\bar{\phi}_1^a\psi_2^b\psi_{3b}$ and $\bar{\phi}_1^a\psi_2^b\psi_3^cf_{abc}$. As an example, consider $(\lambda_1,\lambda_2,\lambda_3)=(1, \half, -\half)$. With an internal SU(3) symmetry, the first variety represents the coupling of gluons to quarks in QCD while the second variety corresponds to the cubic coupling in $\mathcal{N}=4$ Yang-Mills~\cite{BLN}. 
\vskip 0.3cm

\subsection{Scattering amplitudes}
\vskip 0.2cm

Rewrite (\ref{Action}) in momentum space using spinor helicity 
\be
\label{N=4}
\alpha\int \frac{d^4p}{(2\pi)^4}\frac{d^4k}{(2\pi)^4}\frac{d^4l}{(2\pi)^4}(2\pi)^4 \delta^4(p+k+l)\frac{(k_-l-l_-k)^\lambda}{p_-^{\lambda _1}k_-^{\lambda _2+\half}l_-^{\lambda _3+\half}}\tilde{\bar{\phi}}_1(p)\tilde{\psi}_2(k)\tilde{\psi}_3(l)\ ,
\ee
The corresponding amplitude is
\bea
\label{final}
<pk>^{^{(-\lambda_1 +\lambda_2 -\lambda_3)}}\:\:<kl>^{^{(\lambda_1 +\lambda_2 +\lambda_3)}}\:\:<lp>^{^{(-\lambda_1 -\lambda_2 +\lambda_3)}}\ ,
\eea
of the same form as the three-boson case~\cite{AA} and consistent with~\cite{smat,RRM,dixon,bengt}.

\vskip 0.3cm
\section{Theories with supersymmetry: $\mathcal N=4$ Yang-Mills theory}
\vskip 0.2cm
\ndt In this section we discuss this formalism in the context of maximally supersymmetric field theories~\cite{BBB}. The ten-dimensional $\mathcal N=1$ supermultiplet has eight vectors and eight spinors of the little group $SO(8)$. Reduction to four dimensions involves
\be
SO(8)\supset~SO(2)~\times~SO(6)\ ,
\ee
yielding
\be
{\bf 8}^{}_v~=~{\bf 6}^{}_0+{\bf 1}^{}_1+{\bf 1}^{}_{-1}\ ,\qquad 
{\bf 8}^{}_s~=~{\bf 4}^{}_{1/2}+{\bf \bar4}^{}_{-1/2}\ ,
\ee
with $SO(2)$ subscripts. The $(\mathcal N=4, d=4)$ theory contains six scalar fields, one vector field and four spinor fields (and conjugates). Introduce Grassmann variables $\theta^m$ and $\bar\theta_m$ ($m,n, \dots  =1\,,\ldots\, 4$) which transform as the ${\bf 4}$ and $\bf \bar 4$ of $SU(4)$. Their derivatives are
\be
{{\bar \partial}_m}\,~\equiv~\,\frac{\partial}{\partial\,{\theta^m}}\ ;\qquad{\partial^m}\,~\equiv~\,\frac{\partial}{\partial\,{\bar \theta}_m}\ .
\ee
\ndt The physical degrees of freedom of $\mathcal N=4$ Yang-Mills are captured in one superfield~\cite{BLN} 
\bea
\phi\,(y)&=&\frac{1}{ \partial^+}\,A\,(y)\,+\,\frac{i}{\sqrt 2}\,{\theta_{}^m}\,{\theta_{}^n}\,{\nbar C^{}_{mn}}\,(y)\,+\,\frac{1}{12}\,{\theta_{}^m}\,{\theta_{}^n}\,{\theta_{}^p}\,{\theta_{}^q}\,{\epsilon_{mnpq}}\,{\partial^+}\,{\bar A}\,(y)\cr
& &~~~ +~\frac{i}{\partial^+}\,\theta^m_{}\,\bar\chi^{}_m(y)+\frac{\sqrt 2}{6}\theta^m_{}\,\theta^n_{}\,\theta^p_{}\,\epsilon^{}_{mnpq}\,\chi^q_{}(y) \ ,
\eea
where the superfield $\phi$ is not to be confused with the bosonic field, also denoted $\phi$, used in the sections $2-4$. The original eight gauge fields become
\be
A~=~\frac{1}{\sqrt 2}\,(A^{}_1+i\,A^{}_2)\ ,\qquad \bar A~=~\frac{1}{\sqrt 2}\,(A^{}_1-i\,A^{}_2) \ ,
\ee
and the scalars, written as $SU(4)$ bi-spinors 
\bea
C_{}^{m\,4}~=~\frac{1}{\sqrt 2}\,({A^{}_{m+3}}\,+\,i\,{A^{}_{m+6}})\ ,\qquad \nbar C^{}_{m\,4}~=~\frac{1}{\sqrt 2}\,({A^{}_{m+3}}\,-\,i\,{A^{}_{m+6}})\ ,
\eea
for $m\;\neq\,4$. These satisfy
\bea
\label{dual}
{{\nbar C}^{}_{mn}}~=~\,\frac{1}{2}\,{\epsilon^{}_{mnpq}}\,{C_{}^{pq}} \ .
\eea
The fermion fields are $\chi^m$ and $\bar\chi_m$. All fields carry gauge indices and are local in  
\bea
y~=~\,(\,x,\,{\bar x},\,{x^+},\,y^-_{}\equiv {x^-}-\,\frac{i}{\sqrt 2}\,{\theta_{}^m}\,{{\bar \theta}^{}_m}\,)\ .
\eea
The chiral derivatives are
\bea
{d^{\,m}}=-{\partial^m}\,-\,\frac{i}{\sqrt 2}\,{\theta^m}\,{\partial^+}\ ;\qquad{{\bar d}_{\,n}}=\;\;\;{{\bar \partial}_n}\,+\,\frac{i}{\sqrt 2}\,{{\bar \theta}_n}\,{\partial^+}\ .
\eea
The superfield satisfies both the chiral constraint 
\be
{d^{\,m}}\,\phi\,=\,0\ ,
\ee
and the inside-out relation
\be
\label{io}
\bar d_m^{}\,\bar d_n^{}\,\phi~=~\frac{1}{ 2}\,\epsilon_{mnpq}^{}\,d^p_{}\,d^q_{}\,\bar\phi\ ,
\ee
\vskip 0.3cm

\ndt The action for $\mathcal N=4$ Yang-Mills is
\be
\int d^4x\int d^4\theta\,d^4 \bar \theta\,{\cal L}\ ,
\ee
where
\bea
{\cal L}&=&-\bar\phi\,\frac{\Box}{\partial^{+2}}\,\phi
~+\frac{4g}{3}\,f^{abc}_{}\,\Big(\frac{1}{\partial^+_{}}\,\bar\phi^a_{}\,\phi^b_{}\,\bar\partial\,\phi^c_{}+\frac{1}{\partial^+_{}}\,\phi^a_{}\,\bar\phi^b_{}\,\partial\,\bar\phi^c_{}\Big)\cr
&&-g^2f^{abc}_{}\,f^{ade}_{}\Big(\,\frac{1}{\partial^+_{}}(\phi^b\,\partial^+\phi^c)\frac{1}{\partial^+_{}}\,(\bar \phi^d_{}\,\partial^+_{}\,\bar\phi^e)+\frac{1}{2}\,\phi^b_{}\bar\phi^c\,\phi^d_{}\,\bar\phi^e\Big)\ .
\label{ACTION}
\eea
\vskip 0.2cm
\subsection{Generators}
\vskip 0.2cm
\ndt The Lorentz generators, introduced earlier, now accommodate the superspace variables
\be
j~=~x\,\bar\partial-\bar x\,\partial+\frac{1}{ 2}\,(\,{\theta^p}\,{{\bar \partial}_p}\,-\,{{\bar \theta}_p}\,{\partial^p}\,)\,-\lambda\ ,
\ee
with
\be
\lambda~=~\frac{i}{4\sqrt{2}\,\partial^+}\,(\,d^p\,\bar d_p-\bar d_p\,d^p\,)\ .
\ee
$\lambda\,=\,+\,1$ for a chiral superfield. On a chiral superfield, we have
\be
\delta\,\phi~=~i\,\omega\,j\,\phi\ ,\qquad \delta\,\bar \phi~=~-i\,\omega\,j\,\bar \phi\ .
\ee
The other kinematical generators read 
\be
j^+_{}~=~i\, x\,\partial^+_{}\ ,\qquad \bar j^+_{}~=~i\,\bar x\,\partial^+_{}\ ,\qquad  j^{+-}_{}~=~i\,x^-_{}\,\partial^+_{}-\frac{i}{2}\,(\,\theta^p_{}\bar\partial^{}_p+\bar\theta^{}_p\,\partial^p_{}\,)+i\ .
\ee
The boosts now read
\bea
j^-_{}&=&i\,x\,\frac{\partial\bar\partial}{\partial^+_{}} ~-~i\,x^-_{}\,\partial~+~i\,\Big( \theta^p_{}\bar\partial^{}_p\,-\lambda-1\Big)\frac{\partial}{\partial^+_{}}\,\ ,\cr 
\bar j^-_{}&=&i\,\bar x\,\frac{\partial\bar\partial}{\partial^+_{}}~ -~i\,x^-_{}\,\bar\partial~+~ i\,\Big(\bar\theta_p^{}\partial_{}^p+\lambda-1\,\Big)\frac{\bar\partial}{\partial^+_{}}\,\ .
\eea
For further details, we refer the reader to~\cite{PSU}. Half of the supersymmetry generators 
\be
q^{\,m}_{\,+}=-{\partial^m}\,+\,\frac{i}{\sqrt 2}\,{\theta^m}\,{\partial^+}\ ;\qquad{{\bar q}_{\,+\,n}}=\;\;\;{{\bar \partial}_n}\,-\,\frac{i}{\sqrt 2}\,{{\bar \theta}_n}\,{\partial^+}\ ,
\ee
are kinematical while the others are dynamical
\be
{q}^m_{\,-}~\equiv~i\,[\,\bar j^-_{}\,,\,q^{\,m}_{\,+}\,]~=~\frac{\bar\partial}{\partial^+_{}}\, q^{\,m}_{\,+}\ ,\qquad 
{\bar{q}}_{\,-\,m}^{}~\equiv~i\,[\, j^-_{}\,,\,\bar q_{\,+\,m}^{}\,]~=~\frac{\partial}{\partial^+_{}}\, \bar q_{\,+\,m}^{}\ .
\ee
These are ``square-roots" of the Hamiltonian
\be
\{\, {q}^m_{\,-}\,,\,{\bar{q}}_{\,-\,n}^{}\,\}~=~i\,\sqrt{2}\,\delta^{\,m}_{~~n}\,\frac{\partial\bar\partial}{\partial^+_{}}\ .
\ee

\vskip 0.3cm
\subsection{Superconformal algebra}
\vskip 0.2cm
\ndt The $\mathcal N=4$ Yang-Mills theory has a much larger symmstery group than just the usual Poincar\'e symmetry. To build this group: $PSU(2,2\vert\, 4)$, we start with the ``plus" conformal generator~\cite{PSU}
\bea
K^+\,=\,2i\,x\,{\bar x}\,\partial^+\ ,
\eea
which along with $j^{+-}$ yields
\bea
[\,K^{+},\,{p}^{-}\,]=-2i\,D+2i\,j^{+-}\ ,
\eea
the dilatation generator
\bea
D\,=\,i\,\left(\,x^-\partial^+-\,x{\bar \partial}\,-\,{\bar x}\partial-\,\frac{1}{2}\theta\frac{\partial}{\partial \theta}\,-\,\frac{1}{2}{\bar \theta}\frac{\partial}{\partial\bar{\theta}}\,\right)\ .
\eea
Boosting $K^+$ results in
\bea
K\!\!\!&=&\!\!\!i \,[\, j^{-}, K^+\,]=2ix\left(x^-\partial^+-x{\bar \partial}-\theta\frac{\partial}{\partial\,\theta}+\lambda
\right)\!,\\
\bar{K}\!\!\!&=&\!\!\!i\,[\, \bar{j}^{-}, K^+\,]=2i{\bar x}\left(x^-\partial^+-{\bar x}\partial-{\bar \theta}\frac{\partial}{\partial {\bar \theta}}-\lambda
\right)\ .
\eea
The supersymmetry generators now include conformal supersymmetries obtained from
\bea
[\,K^+\,,\,q^m_-\,]\,=\,-\,\sqrt 2\,(\,i\,\sqrt 2\,{\bar x}\,{q^m_+}\,)\,=\,-\,\sqrt 2\,s^m_+\ ,
\eea
and their conjugates, both kinematical. The dynamical conformal  supersymmetries read
\bea
s^m_-\,=\,i\,[\,j^-\,,\,s^m_+\,]=\,i\,\sqrt 2\,\left(\,x^-\,\parp\,-\,x\,\bar \partial\,-\,\theta\,\frac{\partial}{\partial\,\theta}\,
+\lambda+1
\right)\,\frac{1}{\partial^+}\,q^m_+\ ,
\eea
and their conjugates. The dynamical conformal generator $K^{-}$ is 
\be
K^-\,=\,i\,[\,{\bar j}^-\,,\,K\,]\ .
\ee
\vskip 0.3cm
\subsection{Deriving the theory}
\vskip 0.2cm
As before, we begin with an ansatze  for the order $g$ Hamiltonian (we use $g$ instead of $\alpha$ for the coupling constant in supersymmetric Yang-Mills theory). For additional details, we refer the reader to~\cite{BBB, PSU}.
\be
{\delta^g_{p^-}}\,{\phi}~=~-\,i\,g\,{\partial^+}^\mu\,[\,{{\bar \partial}^a}\,{\partial^+}^\rho\,{\phi}\,{{\bar
\partial}^b}\,{\partial^+}^\sigma\,{\phi}\,]\ .
\ee
Exactly like in the non-supersymmetric case, closure of the algebra to order $g$ yields
\be
a\,+\,b\,=\,1\ ,\qquad \mu\,+\,\rho\,+\,\sigma\,=\,0\ .
\ee
The need for a gauge structure function follows as before. The variation that satisfies Poincar\'e invariance to order $g$ is then
\be
{\delta^g_{p^-}}\,{\phi^a}~=~-\,i\,g\,f^{abc}\,\frac{1}{\parp}\,(\,{\bar \partial}\,{\phi^b}\,{\parp}\,{\phi^c})\ .
\ee
In a similar manner, the non-linear boosts are~\cite{PSU}
\bea
\delta^g_{j^-}\,\phi^a~=~-\,x\,\delta^{g}_{p^-}\,\phi^a\,+i\,g\,f^{abc}\,\frac{1}{\partial^+}\,\left\{ \,(\,{\theta}\,
\frac{\partial}{\partial\,\theta}\,-\,1\,)\,{\phi^b}\,{\parp}\,{\phi^c}\,)\,\right\}\ .
\eea
One approach to deriving the Hamiltonian for this theory is to use chirality, dimensional analysis, helicity and elementary commutators to argue that the first order dynamical supersymmetry has the form
\be
\delta^g_{\bar q_-}\,\phi^a~=~-g\,f^{abc}_{}\,\frac{1}{\partial^{+\,(2\nu+1)}}\left\{\,{\bar d}\,\partial_{}^{+\,\nu}\,\phi^b_{}\,\partial^{+\,(\nu+1)}_{}\,\phi^c\right\}\ ,
\ee
where $(d)^4\equiv\epsilon_{mnpq}\,d^md^nd^pd^q$ and $\nu$ is a free parameter. Its conjugate with (\ref {io}) yields
\be
\delta^g_{ q_-}\,\phi^a~=~-g\,f^{abc}_{}\,\frac{(d)^4_{}}{48\,\partial^{+\,(2\nu+3)}}\left\{\,{ d}\,\partial_{}^{+\,\nu}\,\bar\phi^b_{}\,\partial^{+\,(\nu+1)}_{}\,\bar\phi^c\right\}\ ,
\ee
Evaluate 
\be
\{\,\delta^{}_{ q_-^m}\,,\,\delta^{}_{\bar q_{-\,n}}\,\}^g_{}\,\phi^a~=~-\sqrt{2}\,\delta^m_n\,\delta^g_{p^-}\,\phi^a\ ,
\ee
to first order in $g$ to obtain the Hamiltonian
\bea
\label{threen4}
\delta^{}_{p^-}\,\phi^a_{}~=~ -i\,\frac{\partial\bar\partial}{\partial^+}\,\phi^a_{} -\,i\,g\,f^{abc}\,\biggl\{\,\frac {1}{\partial^{+\,(2\nu+1)}}\,(\,{\bar
\partial}\,\partial^{+\,(\nu)}\phi^b\,\partial^{+\,(\nu+1)}\phi^c\,)\nonumber&& \\+\,\frac{(d)^4}{48\,{\parp}^{(2\nu+3)}}\,(\,{\partial}\,\partial^{+\,(\nu)}\,{\bar\phi}^b\,\partial^{+\,(\nu+1)}\,{\bar \phi}^c\,)\,\biggr \}+{\cal O}(g^2)\ .
\eea
The dynamical supersymmetry does not extend to $g^2$~\cite{PSU} and as a consequence, the classical Hamiltonian terminates at order $g^2$. Rather than extend the procedure here to order $g^2$, which is lengthy, we adopt a different approach to arrive at the quartic interaction vertex.
\vskip 0.2cm
\ndt We simply ask that the supersymmetry variations leave the Hamiltonian invariant~\cite{PSU}
\be
\delta^{}_{\bar q_-} H~=~0\ .
\ee
This yields the three conditions
\bea
&&\delta^0_{\bar q_-}\,{{ H}^0}\;=\;0\ , \label{h0}\\
&&\delta^g_{\bar q_-}\,{{ H}^0}\,+\,{\delta^0_{\bar q_-}}\,{{ H}^g}\;=\;0\ , \label{h1}\\
&&{{\delta^g_{\bar q_-}}}\,{{ H}^g}\,+\,{{\delta^0_{\bar q_-}}}\,{{H}^{g^2}}\;=\;0\ ,\label{h2}
\eea
and hence a systematic link to $H^g$ and $H^{g^2}$ from $\delta_{\bar q_-}$ and $H^0$.  The second condition gives
\bea
\delta^g_{\bar q_-}\,{{ H}^0}~=~\delta_{q^{\,-}}^g\,{\biggl \{}\,\int\,\,{{\nbar \phi}^a}\,{\frac {2\,{\partial}{\bar \partial}}{{\parp}^2}}\,{\phi^a}\,{\biggr \}}\ ,
\eea
implying that
\bea
\delta_{q_{\,-}}^g\,{\biggl \{}\,\int\,{{\nbar \phi}^a}\,{\frac {2\,{\partial}{\bar \partial}}{{\parp}^2}}\,{\phi^a}\,{\biggr \}}\,=\,2\,g\,{f^{abc}}\,\int\,{{\nbar \phi}^b}\,d\,{{\nbar \phi}^c}\,{\frac {{\partial}\,{\bar \partial}}{{\parp}^2}}\,{\phi^a}\ .
\eea
so
\be
{\delta^0_{\bar q_-}}\,{{ H}^g}~=~-2\,g\,\int\,{f^{abc}}\,{{\nbar \phi}^b}\,d\,{{\nbar \phi}^c}\,{\frac {{\partial}\,{\bar \partial}}{{\parp}^2}}\,{\phi^a}\ .
\ee
Now consider
\bea
{{\delta^0_{q_{\,-}}}}\,{\biggl \{}\,\,g\,{f^{abc}}\,\int\,{\frac {1}{\parp}}\,{\phi^a}\,{{\nbar \phi}^b}\,{\partial}\,{{\nbar \phi}^c}\,{\biggr \}}\ ,
\eea
which yields two terms~\cite{PSU} which after some manipulations yield
\bea
g\,{f^{abc}}\,\int\,{\frac {1}{\parp}}\,{\phi^a}\,{\frac {\bar \partial}{\parp}}\,d\,{{\nbar \phi}^b}\,{\partial}\,{{\nbar \phi}^c}\,
=\,{\frac {1}{2}}\,{\biggl \{}\,g\,{f^{abc}}\,\int\,{\frac {1}{\parp}}\,{\phi^a}\,{{\nbar \phi}^b}\,{\frac {{\partial}\,{\bar \partial}}{\parp}}\,d\,{{\nbar \phi}^c}\,{\biggr \}}\ .
\eea
Hence the variation
\bea
\delta^0_{q_{\,-}}\,{\biggl \{}\,\,g\,{f^{abc}}\,\int\,{\frac {1}{\parp}}\,{\phi^a}\,{{\nbar \phi}^b}\,{\partial}\,{{\nbar \phi}^c}\,{\biggr \}}\,
=\,\frac{3}{2}\,g\,{f^{abc}}\,\int\,{\frac {1}{\parp}}\,{\phi^a}\,{{\nbar \phi}^b}\,{\frac {{\partial}\,{\bar \partial}}{\parp}}\,d\,{{\nbar \phi}^c}\ ,
\eea
leads to the previously derived cubic vertex in (\ref {threen4}). Next, vary the cubic vertex to obtain the quartic vertex using (\ref{h2}). The cubic vertex involves a part with the transverse derivative $\partial$ and a part with $\bar\partial$. But $H^{g^2}$ does not contain transverse derivatives since it stems from supersymmetries (at order $g$) which do not carry transverse derivatives. Thus
\bea
\label{cons}
\delta^g_{q_-}\,H^g_\partial ~=~0\label{delpart}\ ,
\eea
since $\delta^0_{q_-}$ contains no $\partial$. For supersymmetries to commute with the Hamiltonian then
\be
\label{algerq}
[\,\delta^0_{q^-}\,,\,\delta^{g^2}_{p^-}\,]\,+\,[\,\delta^g_{q^-}\,,\,\delta^g_{p^-}\,]~=~0\ .
\ee
The Hamiltonian contains both $\partial$ and $\bar\partial$ while $\delta^0_{q^-}$ has only $\bar\partial$. Thus one requirement is
\bea
\left.[\,\delta^g_{q^-}\,,\,\delta^g_{p^-}\,]\,\right|_{\mbox{$\partial$}}~=~0\ ,
\eea
which holds as long as the structure functions are antisymmetric and obey the Jacobi identity~\cite{PSU}. The other requirement
\be
\left.[\,\delta^0_{q^-}\,,\,\delta^{g^2}_{p^-}\,]\,+\,[\,\delta^g_{q^-}\,,\,\delta^g_{p^-}\,]\right|_{\mbox{$\bar\partial$}}~=~0\ ,
\ee
leads to $H^{g^2}$
\bea
H_{}^{g^2}
~=~i\int\frac{1}{\partial^+} \bar\phi^a\,\delta^{g^2}_{p^-}\phi^a~=~-\frac{i}{4\sqrt{2}} \int \frac{1}{\partial^+} \bar\phi^a \,\{ \,\delta^g_{q^-}\,,\,\delta^g_{\bar q^-}\,\}\,\phi^a\ .
\eea
The detailed form of the Hamiltonian is presented in~\cite{PSU} and is not relevant here since our aim is to describe the method. Interestingly, this Hamiltonian has the structure of a quadratic form~\cite{PSU}.

\vskip 0.3cm
\section{Theories with supersymmetry: $\mathcal N=8$ supergravity}
\vskip 0.2cm
\ndt Having described $\mathcal N=4$ Yang-Mills, we now move to the other maximally supersymmetric theory in four dimensions, $\mathcal N=8$ supergravity~\cite{n8}. All the fields in $\mathcal N=8$ supergravity are similarly captured by a single superfield~\cite{BLN}. $\theta^m$ now transforms as the $8$ of $SU(8)$.
\bea
\begin{split}
\phi\,(\,y\,)\,=&\,\frac{1}{{\parp}^2}\,h\,(y)\,+\,i\,\theta^m\,\frac{1}{{\parp}^2}\,{\bar \psi}_m\,(y)\,+\,\frac{i}{2}\,\theta^m\,\theta^n\,\frac{1}{\parp}\,{\bar A}_{mn}\,(y)\ , \\
\;&-\,\frac{1}{3!}\,\theta^m\,\theta^n\,\theta^p\,\frac{1}{\parp}\,{\bar \chi}_{mnp}\,(y)\,-\,\frac{1}{4!}\,\theta^m\,\theta^n\,\theta^p\,\theta^q\,{\bar C}_{mnpq}\,(y)\ , \\
\;&+\,\frac{i}{5!}\,\theta^m\,\theta^n\,\theta^p\,\theta^q\,\theta^r\,\epsilon_{mnpqrstu}\,\chi^{stu}\,(y)\ ,\\
\;&+\,\frac{i}{6!}\,\theta^m\,\theta^n\,\theta^p\,\theta^q\,\theta^r\,\theta^s\,\epsilon_{mnpqrstu}\,\parp\,A^{tu}\,(y)\ ,\\
\,&+\,\frac{1}{7!}\,\theta^m\,\theta^n\,\theta^p\,\theta^q\,\theta^r\,\theta^s\,\theta^t\,\epsilon_{mnpqrstu}\,\parp\,\psi^u\,(y)\ ,\\
\,&+\,\frac{4}{8!}\,\theta^m\,\theta^n\,\theta^p\,\theta^q\,\theta^r\,\theta^s\,\theta^t\,\theta^u\,\epsilon_{mnpqrstu}\,{\parp}^2\,{\bar h}\,(y)\ ,
\end{split}
\eea
where the two-component graviton is
\be
\label{graviton}
h\,=\,\frac{1}{\sqrt 2}\,(\,h_{11}\,+\,i\,h_{12}\,)\ ;\qquad {\bar h}\,=\,\frac{1}{\sqrt 2}\,(\,h_{11}\,-\,i\,h_{12}\,)\ .
\ee
\noindent ${\bar \psi}_m$ are the spin-$\frac{3}{2}$ gravitinos, ${\bar A}_{mn}$ the $28$ gauge fields and ${\bar \chi}_{mnp}$, the gauginos. ${\bar C}_{mnpq}$ represents the $70$ scalar fields. Apart from the chiral condition, we have
\bea
\label{io2}
\,{\phi}\,=\,\frac{1}{4}\,\frac{{(d\,)}^8}{{\parp}^4}\,{\bar \phi}\ ,
\eea
\ndt the inside-out relation, with ${(d\,)}^8\,=\,d^1\,d^2\,\ldots\,d^8$. The action to first order in the gravitational coupling constant $\kappa$ reads
\be
\label{n=8}
\beta\,\int\;d^4x\,\int d^8\theta\,d^8 \bar \theta\,{\cal L}\ ,
\ee
where $\beta\,=\,-\,\frac{1}{64}$ and
\bea
\label{one}
{\cal L}&=&-\bar\phi\,\frac{\Box}{\partial^{+4}}\,\phi\,-\,2\,\kappa\,(\,\frac{1}{{\parp}^2}\;{\nbar \phi}\;\;{\bar \partial}\,{\phi}\;{\bar \partial}\,{\phi}+\,\frac{1}{{\parp}^2}\;\phi\,\partial\,{\nbar \phi}\,\partial\,{\nbar \phi})\ ,
\eea
first derived in~\cite{BBB} and subsequently simplified in~\cite{ABR}. 
\vskip 0.2cm
\ndt It is technically very challenging to extend this derivation to order $\kappa^2$ for supergravity. For this reason, we are forced to consider other approaches. As mentioned earlier, the Hamiltonian of $\mathcal =4$ Yang-Mills is a quadratic form~\cite{PSU} and this turns out to be a feature shared by all maximally supersymmetric theories and hence valid for $\mathcal N=8$ supergravity~\cite{ABHS}. 
\vskip 0.2cm
\noindent Using $\mathcal N=4$ Yang-Mills as a guide, the light-cone Hamiltonian for $\mathcal N=8$ supergravity is of the form
\bea
\label{claim}
{\cal H}~=~\frac{1}{4\,\sqrt{2}}\,(\,{\mathcal W}_{\,m}\,,\,{\mathcal W}_{\,m}\,)\ ,
\eea
where
\bea
{\mathcal W}_{\,m}\,=\,{\bar q}_{-\,m}\,\phi\ ,
\eea
and the product is
\be
\label{inner}
(\,\phi\,,\,\xi\,)~\equiv~2i\int d^4\!x\, d^8\theta\,d^8\,{\bar\theta}\;{\bar\phi}\,\frac{1}{{\parp}^3}\xi\ .
\ee
\noindent At lowest order
\bea
\begin{split}
{\cal H}^0\,&=\,\frac{1}{4\,\sqrt{2}}\,(\,{\mathcal W}^0_m\,,\,{\mathcal W}^0_m\,)\ , \\
&=\,\frac{2i}{4\,\sqrt{2}}\,\int d^4\!x\, d^8\theta\,d^8\,{\bar\theta}\;\;q_-^{\,m}\,{\bar \phi}\,\frac{1}{{\parp}^3}\,{\bar q}_{-\,m}\,\phi\ ,
\end{split}
\eea
which can be simplified using (\ref {io2}) to
\bea
{\cal H}^0\,=\,\frac{i}{4\,\sqrt{2}}\,\int d^4\!x\, d^8\theta\,d^8\,{\bar\theta}\;\;{\Big (}\,q_-^{\,m}\,{\bar \phi}\,\frac{1}{{\parp}^3}\,{\bar q}_{-\,m}\,\phi\,+\,\frac{1}{{\parp}^4}\,q_-^{\,m}\,\phi\,\parp\,{\bar q}_{-\,m}\,{\bar \phi}\,{\Big )}\ .
\eea
Putting in the expressions for the supersymmetries we obtain~\cite{ABHS}
\bea
{\cal H}^0\,=\,\int{d^4}x\,{d^8}\theta\,{d^8}{\bar \theta}\,
{\bar \phi}\,\frac{2\,\partial\bar\partial}{{\parp}^4}\,\phi\ ,
\eea
\noindent the appropriate kinetic term in the superspace Hamiltonian of $\mathcal N=8$ supergravity~\cite{BBB}. Moving to order $\kappa$, where the dynamical supersymmetry generators are known~\cite{BBB} we have
\bea
\label{dublew}
{\mathcal W}_m\,=\,\frac{\partial}{\parp} {\bar q}_{+\,m}\,\phi\,+\,\kappa\,\frac{1}{\parp}\,{\Big (}\,{\bar \partial}\,{\bar d}_m\,\phi\,{{\parp}^2}\,\phi\,-\,\parp\,{\bar d}_m\,\phi\,\parp\,{\bar \partial}\,\phi\,{\Big )}\,+\,{\cal O}(\kappa^2)\ ,
\eea
\bea
\label{dublewbar}
{\nbar {\mathcal W}}^m\,=\,\frac{\bar \partial}{\parp}\,q_+^{\,m}\,{\bar \phi}\,+\,\kappa\,\frac{1}{\parp}\,{\Big (}\,\partial\,d^m\,{\bar \phi}\,{{\parp}^2}\,{\bar \phi}\,-\,\parp\,d^m\,{\bar \phi}\,\parp\,\partial\,{\bar \phi}\,{\Big )}\,+\,{\cal O}(\kappa^2)\ .
\eea
which when put into
\bea
\frac{1}{4\,\sqrt{2}}\,(\,{\mathcal W}\,,\,{\mathcal W}\,)\,=\,\frac{2\,i}{4\,\sqrt{2}}\int d^4\!x\, d^8\theta\,d^8\,{\bar\theta}\;{\nbar {\mathcal W}}\,\frac{1}{{\parp}^3}\,{\mathcal W}\ .
\eea
yield~\cite{ABHS}
\bea
\int\;d^4x\,d^8\theta\,d^8{\bar \theta}\;\;2\,\kappa\,{\Big (}\,\frac{1}{{\parp}^2}\;{\nbar \phi}\;\;{\bar \partial}\,{\phi}\;{\bar \partial}\,{\phi}+\,\frac{1}{{\parp}^2}\;\phi\,\partial\,{\nbar \phi}\,\partial\,{\nbar \phi}{\Big )}\ ,
\eea
which is the cubic interaction vertex - and matches that derived by gauge fixing the covariant theory in (\ref {one}).
\vskip 0.3cm
\ndt This quadratic form structure leads us naturally to a quartic interaction vertex~\cite{ABHS} but the result is too cumbersome to be useful in calculations. There is a more recent coherent state strategy that seems to make more sense but again its use in explicit computations is unclear at this point. The way forward here appears to be through the exceptional symmetry in the theory. In particular, using the $E_{7(7)}$ symmetry in $\mathcal N=8$ supergravity in conjunction with the quadratic form approach yields a much simpler quartic interaction vertex~\cite{BR}.
\vskip 0.3cm

\ndt As far as $\mathcal N=8$ supergravity is concerned, it seems fairly clear that there are two distinct approaches to building the theory. The first is to use the superPoincar\'e algebra exclusively while the other is to follow the exceptional symmetry in the theory~\cite{ABM}. Clearly a field redefinition should relate the results of these two approaches. 
\vskip 0.2cm
\ndt It is also interesting to note that this valuable exceptional symmetry grows under dimensional reduction~\cite{Exc}. In $d=3$ it becomes an $E_{8(8)}$ symmetry, an $E_9$ symmetry in $d=2$, an $E_{10}$ symmetry in $d=1$ and an $E_{11}$ symmetry in $d=0$. It seems possible that some or all of these higher symmetry groups could be present in the eleven-dimensional theory itself. Their link to clarifying the finiteness issue~\cite{fin} relating to the $\mathcal N=8$ model is certainly worth examining further.

\vskip .4cm
\begin{center}
* ~ * ~ *
\end{center}

\ndt 
There are a number of open questions worth exploring. First, can this light-cone symmetry-based approach provide a Lagrangian origin to the Vasiliev equations of motion in $AdS_4$? Second, can we derive consistent quartic interaction vertices in flat spacetime? If not, where exactly does the procedure fail and how does this fit in with the exisiting no-go literature pertaining to higher spins in flat spacetime. Third, can this approach offer hints regarding the missing ingredients for the $(\mathcal N=2, 0)$ theory in $d=6$? In particular, since we know the six-dimensional supercofrmal algebra in light-cone gauge can we close commutators and zero in on what structures are missing~\cite{Ramond}?
\vskip 0.5cm
\ndt {\it {Acknowledgments}}
\vskip 0.3cm

\ndt I thank the organizers of the “Higher Spin Gauge Theories” workshop at NTU and the Institute of Advanced Studies, NTU for the hospitality. I am grateful to Y. S. Akshay, Lars Brink, Mahendra Mali, Sucheta Majumdar, Pierre Ramond and Hidehiko Shimada for discussions.

\end{document}